\documentclass[final]{svjour3}
\usepackage{graphicx}
\usepackage{amssymb}
\usepackage{multirow}
\usepackage{mathptmx}
\usepackage[numbers]{natbib}
\usepackage{subfigure}
\makeatletter
\journalname{Journal of Low Temperature Physics}

\bibpunct{}{}{,}{s}{}{,}

\begin{document}

\newcommand{\hdblarrow}{H\makebox[0.9ex][l]{$\downdownarrows$}-}
\title{The POLARBEAR-2 and Simons Array Focal Plane Fabrication Status}

\author{Benjamin Westbrook \and The POLARBEAR Collaboration}


\author{
B.~Westbrook${^i}$\and
P.~A.~R.~Ade${^b}$\and
M.~Aguilar${^c}$\and
Y.~Akiba${^{d,e}}$\and
K.~Arnold${^a}$\and
C.~Baccigalupi${^f}$\and
D.~Barron${^g}$\and
D.~Beck${^h}$\and
S.~Beckman${^i}$\and
A.~N.~Bender${^{j,k}}$\and
F.~Bianchini${^l}$\and
D.~Boettger${^m}$\and
J.~Borrill${^{g,n}}$\and
S.~Chapman${^o}$\and
Y.~Chinone${^{i,p}}$\and
G.~Coppi${^q}$\and
K.~Crowley${^a}$\and
A.~Cukierman${^i}$\and
T.~de Haan${^r}$\and
R.~D\"unner${^m}$\and
M.~Dobbs${^s}$\and
T.~Elleflot${^a}$\and
J.~Errard${^h}$\and
G.~Fabbian${^t}$\and
S.~M.~Feeney${^u}$\and
C.~Feng${^v}$\and
G.~Fuller${^a}$\and
N.~Galitzki${^a}$\and
A.~Gilbert${^s}$\and
N.~Goeckner-Wald${^i}$\and
J.~Groh${^i}$\and
N.~W.~Halverson${^{w,x,y}}$\and
T.~Hamada${^{e,z}}$\and
M.~Hasegawa${^{e,d}}$\and
M.~Hazumi${^{d,e,p,aa}}$\and
C.~A.~Hill${^{i,r}}$\and
W.~Holzapfel${^{i}}$\and
L.~Howe${^{a}}$\and
Y.~Inoue${^{e,bb}}$\and
G.~Jaehnig${^{w,y}}$\and
A.~Jaffe${^{cc}}$\and
O.~Jeong${^i}$\and
D.~Kaneko${^p}$\and
N.~Katayama${^p}$\and
B.~Keating${^a}$\and
R.~Keskitalo${^{g,n}}$\and
T.~Kisner${^{g,n}}$\and
N.~Krachmalnicoff${^f}$\and
A.~Kusaka${^{r,dd}}$\and
M.~Le Jeune${^h}$\and
A.~T.~Lee${^{i,r,ee}}$\and
D.~Leon${^{a}}$\and
E.~Linder${^{g,r}}$\and
L.~Lowry${^{a}}$\and
A.~Madurowicz${^{i,r}}$\and
D.~Mak${^{cc}}$\and
F.~Matsuda${^{a}}$\and
A.~May${^q}$\and
N.~J.~Miller${^{ff}}$\and
Y.~Minami${^e}$\and
J.~Montgomery${^s}$\and
M.~Navaroli${^a}$\and
H.~Nishino${^e}$\and
J.~Peloton${^{gg}}$\and
A.~Pham${^l}$\and
L.~Piccirillo${^q}$\and
D.~Plambeck${^{ee}}$\and
D.~Poletti${^f}$\and
G.~Puglisi${^f}$\and
C.~Raum${^i}$\and
G.~Rebeiz${^a}$\and
C.~L.~Reichardt${^l}$\and
P.~L.~Richards${^i}$\and
H.~Roberts${^{w,y}}$\and
C.~Ross${^{hh}}$\and
K.~M.~Rotermund${^{hh}}$\and
Y.~Segawa${^{d,e}}$\and
B.~Sherwin${^{r,ii,jj}}$\and
M.~Silva-Feaver${^{a}}$\and
P.~Siritanasak${^{a}}$\and
R.~Stompor${^h}$\and
A.~Suzuki${^{i,ee}}$\and
O.~Tajima${^{d,e}}$\and
S.~Takakura${^{e,kk}}$\and
S.~Takatori${^{d,e}}$\and
D.~Tanabe${^{d,e}}$\and
R.~Tat${^{i,r}}$\and
G.~P.~Teply${^{a}}$\and
A.~Tikhomirov${^{hh}}$\and
T.~Tomaru${^e}$\and
C.~Tsai${^a}$\and
N.~Whitehorn${^i}$\and
A.~Zahn${^a}$}

\institute{
\email{bwestbrook@berkeley.edu}\\
${^a}$Department of Physics, University of California, San Diego, CA 92093-0424, USA\and
${^b}$School of Physics and Astronomy, Cardiff University, Cardiff CF10 3XQ, United Kingdom\and
${^c}$Departamento de Fisica, FCFM, Universidad de Chile, Blanco Encalada 2008, Santiago, Chile\and
${^d}$The Graduate University for Advanced Studies (SOKENDAI), Miura District, Kanagawa 240-0115, Hayama, Japan\and
${^e}$High Energy Accelerator Research Organization (KEK), Tsukuba, Ibaraki 305-0801, Japan\and
${^f}$International School for Advanced Studies (SISSA), Via Bonomea 265, 34136, Trieste, Italy\and
${^g}$Space Sciences Laboratory, University of California, Berkeley, CA 94720, USA\and
${^h}$AstroParticule et Cosmologie (APC), Univ Paris Diderot, CNRS/IN2P3, CEA/Irfu, Obs de Paris, Sorbonne Paris Cite, France\and
${^i}$Department of Physics, University of California, Berkeley, CA 94720, USA\and
${^j}$Argonne National Laboratory, High-Energy Physics Division, 9700 S. Cass Avenue, Argonne, IL, USA 60439\and
${^k}$Kavli Institute for Cosmological Physics, University of Chicago, 5640 South Ellis Avenue, Chicago, IL 60637\and
${^l}$School of Physics, University of Melbourne, Parkville, VIC 3010, Australia\and
${^m}$Instituto de Astrofisica and Centro de Astro-Ingenieria, Facultad de Fisica, Pontificia Universidad Catolica de Chile, Av. Vicuna Mackenna 4860, 7820436 Macul, Santiago, Chile\and
${^n}$Computational Cosmology Center, Lawrence Berkeley National Laboratory, Berkeley, CA 94720, USA\and
${^o}$Department of Physics and Atmospheric Science, Dalhousie University, Halifax, NS, B3H 4R2, Canada\and
${^p}$Kavli Institute for the Physics and Mathematics of the Universe (Kavli IPMU, WPI), UTIAS, The University of Tokyo, Kashiwa, Chiba 277-8583, Japan\and
${^q}$The University of Manchester, Manchester M13 9PL, United Kingdom\and
${^r}$Physics Division, Lawrence Berkeley National Laboratory, Berkeley, CA 94720, USA\and
${^s}$Physics Department, McGill University, Montreal, QC H3A 0G4, Canada\and
${^t}$Institut d'Astrophysique Spatiale, CNRS (UMR 8617), Univ. Paris-Sud, Universite Paris-Saclay, bat. 121, 91405 Orsay, France\and
${^u}$Center for Computational Astrophysics, Flatiron Institute, 162 5th Avenue, New York, NY 10010, USA\and
${^v}$Department of Physics and Astronomy, University of California, Irvine, CA 92697-4575, USA\and
${^w}$Center for Astrophysics and Space Astronomy, University of Colorado, Boulder, CO 80309, USA\and
${^x}$Department of Astrophysical and Planetary Sciences, University of Colorado, Boulder, CO 80309, USA\and
${^y}$Department of Physics, University of Colorado, Boulder, CO 80309, USA\and
${^z}$Astronomical Institute, Graduate School of Science, Tohoku University, Sendai, 980-8578, Japan\and
${^{aa}}$Institute of Space and Astronautical Science (ISAS), Japan Aerospace Exploration Agency (JAXA), Sagamihara, Kanagawa 252-0222, Japan\and
${^{bb}}$Institute of Physics, Academia Sinica, 128, Sec.2, Academia Road, Nankang, Taiwan\and
${^{cc}}$Department of Physics, Imperial College London, London SW7 2AZ, United Kingdom\and
${^{dd}}$Department of Physics, The University of Tokyo, Tokyo 113-0033, Japan\and
${^{ee}}$Radio Astronomy Laboratory, University of California, Berkeley, CA 94720, USA\and
${^{ff}}$Observational Cosmology Laboratory, Code 665, NASA Goddard Space Flight Center, Greenbelt, MD 20771, USA\and
${^{gg}}$Department of Physics \& Astronomy, University of Sussex, Brighton BN1 9QH, UK\and
${^{hh}}$Department of Physics and Atmospheric Science, Dalhousie University, Halifax, NS, B3H 4R2, Canada\and
${^{ii}}$DAMTP, University of Cambridge, Cambridge CB3 0WA, UK\and
${^{jj}}$Kavli Institute for Cosmology Cambridge, Cambridge CB3 OHA, UK\and
${^{kk}}$Osaka University, Toyonaka, Osaka 560-0043, Japan\\
}

\maketitle

\begin{abstract}

We present on the status of POLARBEAR-2 A (PB2-A) focal plane fabrication.  The PB2-A is the first of three telescopes in the Simon Array (SA), which is an array of three cosmic microwave background (CMB) polarization sensitive telescopes located at the POLARBEAR (PB) site in Northern Chile.  As the successor to the PB experiment, each telescope and receiver combination is named as PB2-A, PB2-B, and PB2-C.  PB2-A and -B will have nearly identical receivers operating at 90 and 150 GHz while PB2-C will house a receiver operating at 220 and 270 GHz.  Each receiver contains a focal plane consisting of seven close-hex packed lenslet coupled sinuous antenna transition edge sensor bolometer arrays.   Each array contains 271 di-chroic optical pixels each of which have four TES bolometers for a total of 7588 detectors per receiver. We have produced a set of two types of candidate arrays for PB2-A.  The first we call Version 11 (V11) and uses a silicon oxide (SiOx) for the transmission lines and cross-over process for orthogonal polarizations.  The second we call Version 13 (V13) and uses silicon nitride (SiNx) for the transmission lines and cross-under process for orthogonal polarizations.  We have produced enough of each type of array to fully populate the focal plane of the PB2-A receiver.  The average wirebond yield for V11 and V13 arrays is 93.2\% and 95.6\%  respectively.   The V11 arrays had a superconducting transition temperature ($T_c$) of 452 $\pm$ 15 mK, a normal resistance ($R_n$) of 1.25 $\pm$ 0.20 $\Omega$, and saturations powers of 5.2 $\pm$ 1.0 $pW$ and 13 $\pm$ 1.2 $pW$ for the 90 and 150 GHz bands respectively.  The V13 arrays had a superconducting transition temperature ($T_c$) of 456 $\pm$ 6 mK, a normal resistance ($R_n$) of 1.1 $\pm$ 0.2 $\Omega$, and saturations powers of 10.8 $\pm$ 1.8 $pW$ and 22.9 $\pm$ 2.6 $pW$ for the 90 and 150 GHz bands respectively.   Production and charcterization of arrays for PB2-B are ongoing and are expected to be completed by the summer of 2018.  We have fabricated the first three candidate arrays for PB2-C but do not have any characterization results to present at this time.  


\end{abstract}

\vspace{-25pt}
\section{Introduction}

\vspace{-10pt}
Currently there are many experiements attempting to measure the polarization the CMB to uprecedented precision.  Many of these  polarimeters currently being built, deployed, and/or operated\cite{Henderson_AdvancedAct_2016, Wu_BICEP3_2016, Suzuki_SA_2016, Benson_SPT3G_2014, Gualtieri_SPIDER_2017} while others are in the design and funding stages\cite{Suzuki_LB_2018, CMBS4_2017, Hill_SO_2018}. This work focuses on the fabrication and characterization of the focal plane for the PB2-A telescope of the Simons Array.\cite{Inoue_PB2_2016} 

We fabricate the arrays at the Marvell Nanofabrication Facility (MNF) at UC Berkeley using a series of thin film deposition, photo-lithographic, and etching techniques.\cite{Westbrook_Sinuous_2016}  The PB2-A focal plane requires seven 90/150 dichroic arrays that meet the peformance requirements outlined in Table~\ref{table:SA_Requirements}.

\vspace{-5pt}
\begin{table}[htbp]
\begin{center}
\begin{tabular}{| c | c | c | c |}
 \multicolumn{1}{c}{Band} & \multicolumn{1}{c}{ Specification} &  \multicolumn{1}{c}{Version 11} &   \multicolumn{1}{c}{Version 13} \\
\hline
\multirow{3}{*}{90} & Band Center (GHz) & 89.5 $\pm$ 4.5 & 89.5 $\pm$ 4.5  \\
& Fractional BW (GHz) & 0.324 $\pm$ 0.032 & 0.324 $\pm$ 0.032  \\
& Saturation Power ($pW$) & 5-7 & 7-9  \\
\hline
\multirow{3}{*}{150} & Band Center (GHz) & 147.5 $\pm$ 7.4 & 147.5 $\pm$ 7.4\\
& Fractional BW (GHz) & 0.260 $\pm$ 0.026 & 0.260 $\pm$ 0.026 \\
& Saturation Power ($pW$) & 9-15 & 17-24  \\
\hline
\multirow{4}{*}{All} & Warm Yield (\%) & \multicolumn{2}{c |}{$>$90\%}   \\
& Normal Resistance $R_n$ ($\Omega$) & \multicolumn{2}{c |}{1.2 $\pm$ 0.3}   \\
& Operating Resistance $R_n$ ($\Omega$) &\multicolumn{2}{c |}{ 0.8 $\pm$ 0.2}  \\
& Transition Temperature $T_c$ (mK) & \multicolumn{2}{c |}{420-470}  \\
& Intrinsic Time Constant $\tau_0$ (ms) & \multicolumn{2}{c |}{10-25}  \\
\hline
\end{tabular}
\caption{A table breaking down the requirements for the PB2-A bolometer arrays.  The $R_{n}$ requirement is set by the DfMux readout and is the same for all bands.  The $P_{sat}$ requirement is set by optical loading and sensitivity calculations for each band and is therefore different for each band. The $T_c$ requirement is set by optimizing phonon-noise for a given saturation power.  In practice the range is quite broad as we can tune the saturation power ($P_{sat}$) of the detectors by modulating bolometer geometry given a $T_c$ with a minimal noise hit.  The difference in satuartion power targets for V11 and V13 comes from the fact that we expect greater optical efficiency with the V13 design and are quoted as ranges to accommodate multiple HWP locations and variance in weather.  The PB2-A bands are choosen to maximize sensitivity for the CMB within the atmospheric windows avaialble in Chile while avoiding the carbon monoxide (CO) lines at 110 and 115 GHz.  The intrinsic time constant $\tau_0$ is set by the readout bandwidth and the rotation rate of the HWP.}\label{table:SA_Requirements}
\end{center}
\end{table}

\begin{figure}[htbp]
\begin{center}
\includegraphics[width=\textwidth]{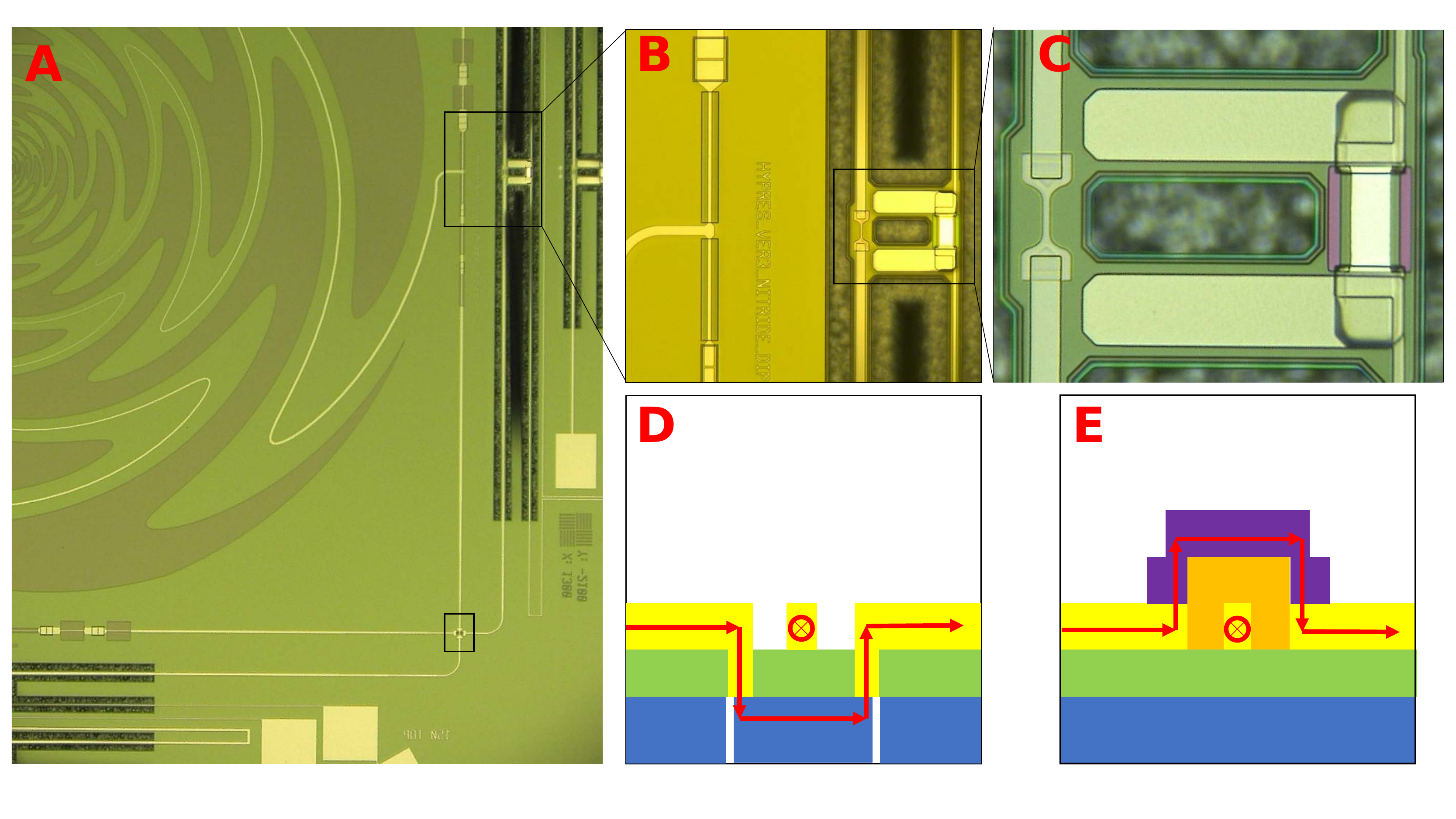}
\caption{A series of photographs detailing a PB2-A pixel. Radiation excites a broadband sinuous antenna, which is coupled to a pair of orthogonal microstrip transmission lines (panels A and B).  Either ``cross overs'' in V11 (panel D) or ``cross unders'' in V13 (panel E) electrically isolate each polarization.  The radiation is carried to a pair of lumped element filters by the niobium (Nb) mictrostrip, which separates the incoming radiation into bands (panels B and C). The power from each band is then terminated on a titanium (Ti) resistor and converted to heat and coupled to a comparatively large pair of palaldium (Pd) thermal ballasts (panel C).  The temperature fluctuations are measured using digital frequency division multplexing (DfMUX) to operate the Aluminum Manganese (AlMn) TES's in a negative electrothermal feedback (panel C).\cite{DfMUX_Dobbs_2012}}\label{fig:pixel_structure}
\end{center}
\end{figure}

\vspace{-35pt}
Throughout this process we have tried 13 different design iterations which we call versions.   Initially we focused on SiOx as the dielectric material and implemented this design using a ``cross over" structure to electrically isolate orthogonal polarizations (see Fig.~\ref{fig:pixel_structure}).  We converged on a reliable fabrication recipe on the eleventh iteration and will refer to these arrays from now as V11 arrays.  After fabricating enough V11 arrays to deploy, we PB2-A, we began work on arrays which use SiNx as the dielectric for the microstrip (see Fig.~\ref{fig:pixel_structure}).  The loss tangent for SiNx is about a factor $\sim$10 lower than SiOx, which translates to an increase in optical efficiency of the each pixel by a factor of $\sim$1.5-2.  We have developed a process using SiNx as the dielectric and a ``cross under"  for the orthogonal polarizations, which we call Version 13 (V13).  The V13 process eliminates two deposition and three lithograhic steps compared to V11 process which translates to a 3\% increase in warm fabrication yield.  We discuss the current production and characterization status of both V11 and V13 arrays.   

\vspace{-20pt}
\section{The POLARBEAR-2 A Focal Plane}

\begin{figure}[ht!]
\begin{center}
\includegraphics[width=\textwidth]{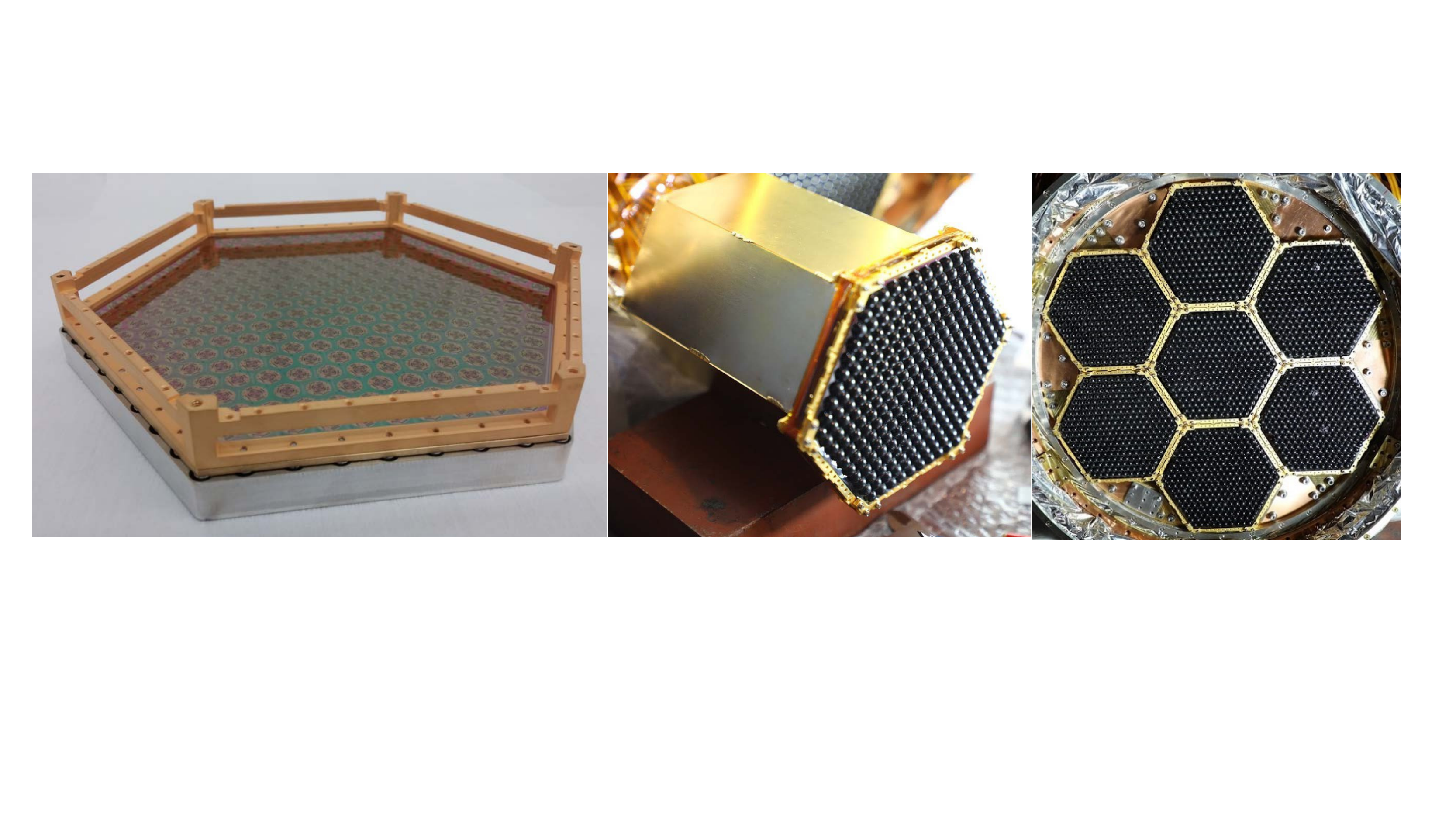}
\end{center}
\vspace{-0.75in}
\caption{Photographs of a PB2-A candidate array at three different stages of assembly for integration into the main receiver.  The bolometer array and lenslet array (behind the device array) are aligned and mounted to a hex-shaped gold-plated invar structure (left).  We chose invar to minimizes differential thermal contraction between the support structure and the arrays.  The gold plating mazimizes thermal conductivity while having low emmissivity.  The module is assembled with cryogenic readout hardware incased in a hexangonal gold plated tube, which provides RF and mechanical shielding for the readout electronics (middle).  Wirebonds (not shown) from bond pads at the edge of the arrays connect the to cold readout components which cannot be seen in this photograph.  Seven candidate arrays installed into the PB2-A focal plane (right).}\label{fig:Focal_Plane_Photos}
\end{figure}

\vspace{-10pt}
The PB2-A focal plane contains of seven close-hex packed broadband sinous antenna TES bolometer arrays.   Each array has 271 polarization sensitive pixels with two TES bolometers per band and polarization (four per pixel) for a total of 7,588 TES bolometers.   The bolometer arrays are read out with 40X digital frequency division multplexing (DfMUX) via cold readout hardware mounted to the back of each module. \cite{DfMUX_Dobbs_2012, Elleflot_PB2Readout_2018} Each pixel is optically coupled to the telescope with hemispherical silicon lenslets as shown in the middle and right panels of Fig.~\ref{fig:Focal_Plane_Photos}.   

\vspace{-20pt}
\subsection{Pixel Design}

\vspace{-10pt}
For the purposes of this paper, each pixel contains two basic stuctures: the RF circuit and the bolometers and are described in detail in Fig.~\ref{fig:pixel_structure}. 

\vspace{-15pt}
\subsubsection{RF circuit:}  

\vspace{-10pt}
The RF circuit is composed of a polarization-sesnsitive broadband sinuous antenna, a superconducting microstrip Niobium (Nb) transmission lines, two 3-pole Chebyshev filters, and a titanium (Ti) termination resistor.  Each pixel is coupled to a hemispherical lenslet which focuses light from the telescope on to each of the sinusous antennae in the array.  The Nb microstrips (1 for each polarization axis) carry the radiation to the di-chroic filters that split the radiation into two bands each with $\sim$30\% fractional bandwidth as shown in panels A and B of Fig.~\ref{fig:pixel_structure}.  The transmission lines are contructed with a Nb ground plane coated with a dielectric material (SiOx or SiNx) and Nb microtrip line patterned on top of that as shown in all panels of Fig.~\ref{fig:pixel_structure}). The RF circuit requires that orthogonal transmission lines cross one another while being electrically isolated.  This is acheived with either building a ``cross over" or a ``cross under," which allow each polarization to be sensed by separate bolometers as shown in panels D and E of Fig.~\ref{fig:pixel_structure}.     We designate bolometers to be ``top" (T) or ``bottom"(B) bolometers if their microstrip forms the top or bottom portion of the cross over/under. The power from each band and polarization is converted to heat by lumped element titanium (Ti) located at the center of bolometers shown in panel C of Fig.~\ref{fig:pixel_structure}.

\vspace{-15pt}
\subsubsection{Bolometers:}  

\vspace{-10pt}
Fluctuations in power delivered to the Ti resistor are thermally coupled to and sensed by a TES via high-heat capacity thermal ballasts made of Palladium (Pd) as seen in panel C of Fig.~\ref{fig:pixel_structure}.  Superconducting Nb leads provide voltage bias for the digital frequency division multiplexing (DfMux) readout by routing connections for each TES to bond pads at the edges of the array.\cite{Elleflot_PB2Readout_2018}  The bolometers are four-legged floating structures that provide thermal isolation from the 250~mK bath.  Two of the legs carry the filtered power from each band and polarization along microstrip to Ti termination resistors while the other two legs hold Nb leads which provide the AlMn TES with bias for readout.  This design allows for straight forward modication of the bolometric properites by simply scaling the geometry of each structure.  For example, we make modifications to the $R_n$ of the array by simply changing the number of squares for the TES and to the $P_{sat}$ simply by changing the length of the legs of the bolometer structure.  The time constant of the bolometer is tuned by changing the volume (foot print and thickness) of the Pd thermal ballast.

\vspace{-25pt}
\section{Characterization Status}

\vspace{-10pt}
The SA focal planes are fabricated at UC Berkeley by two dedicated fabrication engineers.  We have demonstrated that we can fabricate 1 array approximately every 2 weeks per fabrication engineer on average when accounting for machine down time.   For both the V11 and V13 processes, we yield complete $\sim$90\% of the arrays that we begin processing on.  The 10\% loss comes from unexpected tool behavoir and/or handling errors and are not a flaw of the fabrication process.  Once a array is completed with no reported issues from the fabrication engineer, we begin to characterize the array.  To date, we have finished the fabriation of seven V11 arrays and eight V13 arrays and characterization is ongoing, which will determine the final configuration of the PB2-A plane.  The target specficiations of the arrays are set by sensitivity requirements of PB2-A and are shown in Table~\ref{table:SA_Requirements}.  

\vspace{-20pt}
\subsection{Room Temperature Characterization}

\vspace{-10pt}
Room temperature characterizations starts with process monitoring and visual inspection at each step of the fabrication process to check for arrays obvious fabrication errors.  Any individual process that have visual defects and/or deviate from our process specficiations require further inspection to determine if they should be discarded, reworked, or are suitable for futher processing. Once we have completed an array in a nominal manner, we measure linewidth of Nb microstrip, the Ti resistor, and the TES as well as the thickness of the microstip dielectric layer at 9 points across the array to ensure we have good fidelity with respect to the design.   We find that linewidth varies by less than 5\% across the wafer and dielectric thickess varies by less than 3\% across the wafer. We also take electrical mesurements of intrapixel test structures in these nine locations to measure the room temperature resistance of Ti resistor and TES as well as cross over/under symmetry and isolation. 

Arrays which pass all of these criteria are then prepared for cryogenic tests by wirebonding the detectors and mating them with a lenslet array and the necessary cold readout hardware.\cite{Elleflot_PB2Readout_2018}   The resistance of the leads is proportional to length and therefore the resistance decreases with radial distance from the center, which is used to define the expected resistance of a given bolometer and its Nb leads.  These resistance checks are used to screen the Nb leads for shorts between pairs of leads, shorts to nearest neighbors, and shorts to the ground plane.  We define wirebond yield to be all bolometers which do not have any shorts and have the expected lead resistance divided by 1140, which includes all the possible channels we have readout for.  Table~\ref{table:Wirebond_yield} shows an example resistance heat map for a V13 array and tabluates yield for 15 candidate V11 and V13 arrays.  We find that problematic shorts and opens are randomly distributed across the array and are consistent with particles and pin hole defects in the photoresist during processing creating unwanted holes and/or bridges in our patterned structures.  At this time we do not report the end to end cryogenic yield. 

\begin{table}[htbp]
\begin{center}
\begin{tabular}{| c | c | c |}
\hline
& & \\
  \multirow{10}{*}{ \includegraphics[width=0.42\textwidth]{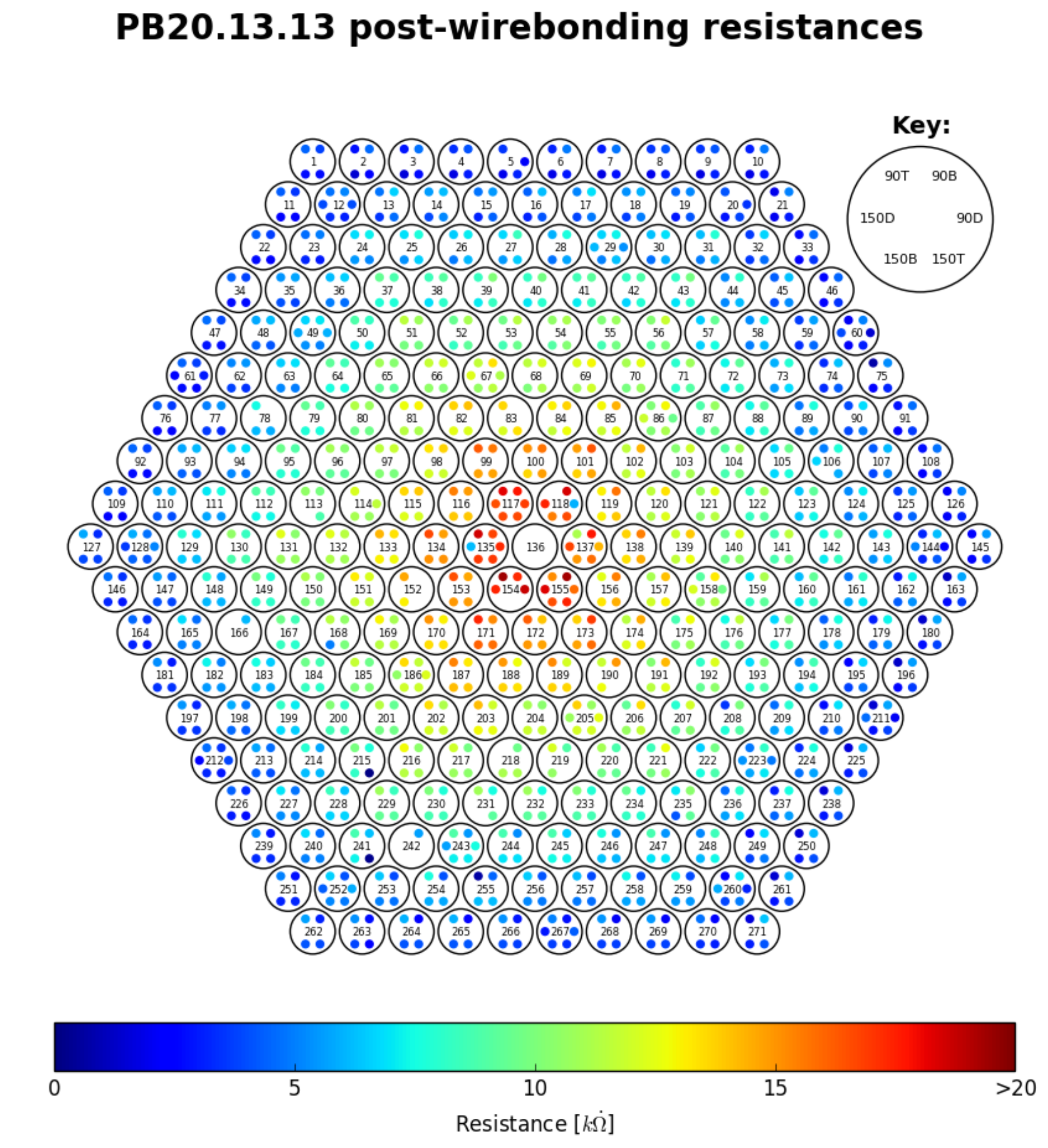}}& Array Name & Wirebond Yield (\%) \\
 & PB2-11-01 & 92.5 \\
 & PB2-11-02 & 95.7 \\
 & PB2-11-03 & 93.5 \\
 & PB2-11-04 & 88.2 \\ 
 & PB2-11-05 & 95.0 \\
 & PB2-11-06 & 90.2 \\
 & PB2-11-08 & 94.0 \\
 & \bf V11 Avg & \bf 92.5 \\
 & PB2-13-09 & 96.5\\
 & PB2-13-10 & 96.6\\
 & PB2-13-11 & 92.5\\
 & PB2-13-12 & 95.3\\
 & PB2-13-13 & 97.4 \\ 
 & PB2-13-14 & 96.1 \\
 & PB2-13-15 & 95.2 \\
 & PB2-13-16 & 96.5  \\
 & \bf V13 Avg & \bf 96.3 \\
 & & \\
\hline
\end{tabular}
\caption{Summary of the warm wired bond yield characterization data for 15 candidate arrays for PB2-A and PB2-B, which is used to define wirebond yield.  V13 arrays have better yield than V11 due to the reduction in steps required to make them as the cross-under process requires three fewer steps compared to the cross-over process. }\label{table:Wirebond_yield}
\end{center}
\end{table}

Any arrays that do not have sufficient wirebond yield at room temperature and/or have visually identifiable or known defects are discarded.  However, we find that arrays that pass visual inspection also meet the wirebond yield requirement of greater than 90\%. The V11 arrays have an average warm yield of 92.7\% and the V13 arrays have an average warm yield of 95.6\% as shown in Table~\ref{table:Wirebond_yield}. 

\vspace{-20pt}
\subsection{Cryogenic Characterization}

\vspace{-10pt}
Arrays that pass warm characterization are then cooled down in one of the test beds available to our collaboration to validate the cryogenic properties of the arrays.\cite{Elleflot_PB2Readout_2018} At the moment, complete characterization of our candidate arrays is on going, so we only report on a subset of all bolometer properties for a subset of our candidate arrays.   We split cryogenic characterization into two categories: dark and optical.  Since the majority of our test beds are dark cryostats we provide comparatively more data on dark bolometric properties.

\vspace{-15pt}
\subsubsection{Dark Characteriztation}

\begin{figure}[htbp]
\begin{center}
\begin{tabular}{ | c | c  | c  |}
\hline
& & \\
Parameter & V11 & PB20.13.11 \\
\hline
\hline
& & \\
$R_n$ ($\Omega$)  &\includegraphics[width=0.3\textwidth]{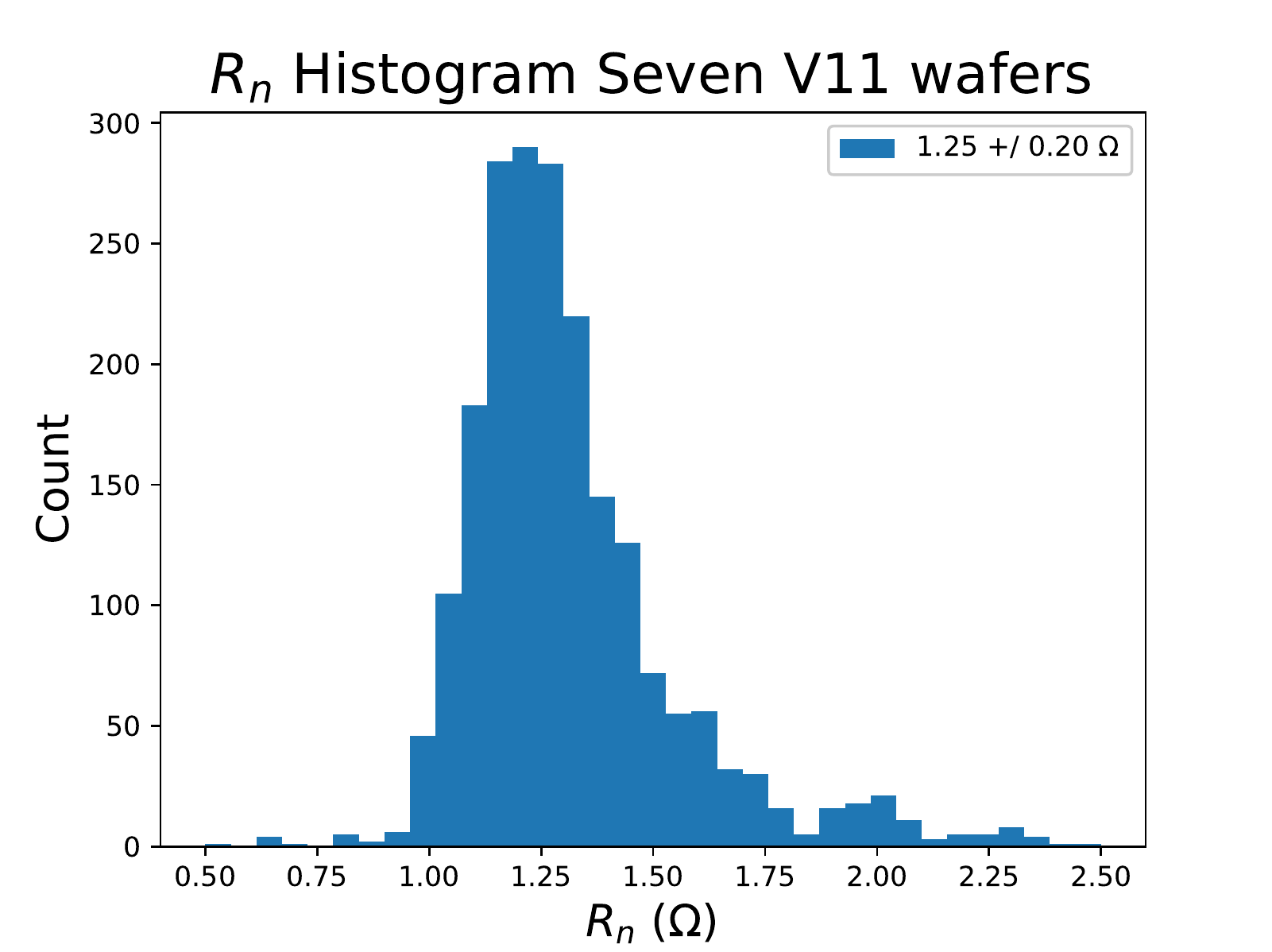} & \includegraphics[width=0.3\textwidth]{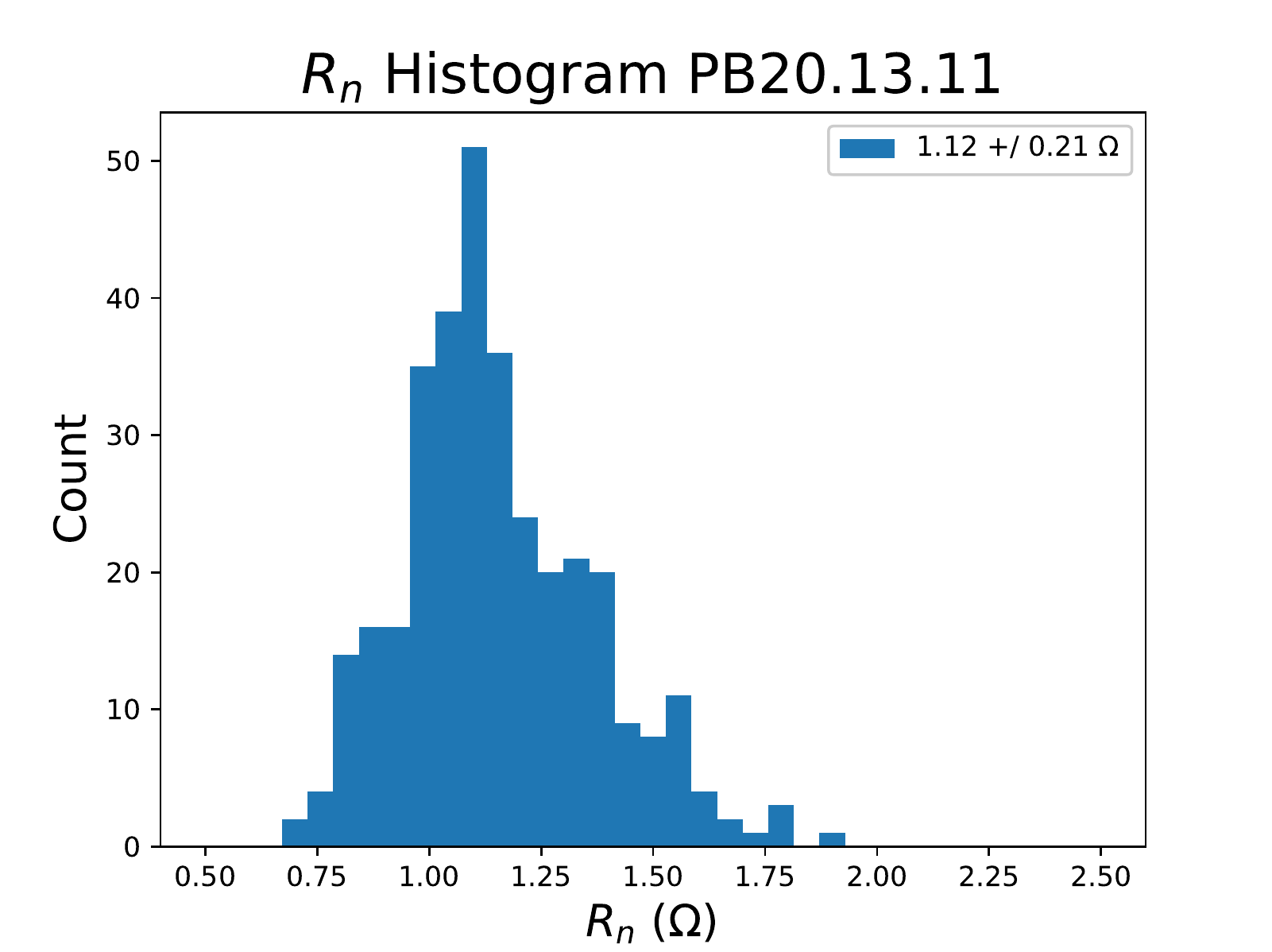} \\
\hline
\hline
& & \\
 $T_c$ ($mK$)&  \includegraphics[width=0.3\textwidth]{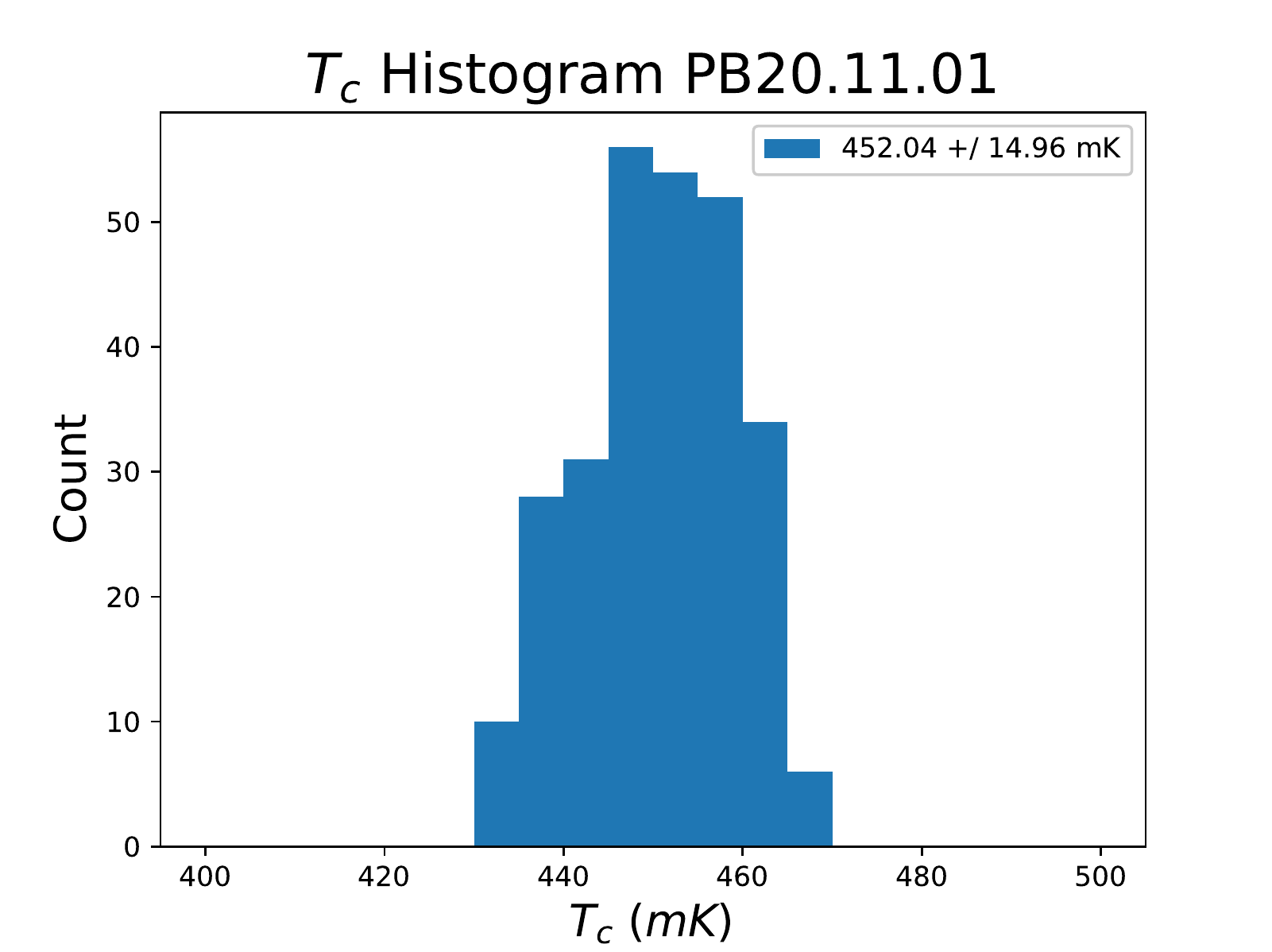} &  \includegraphics[width=0.3\textwidth]{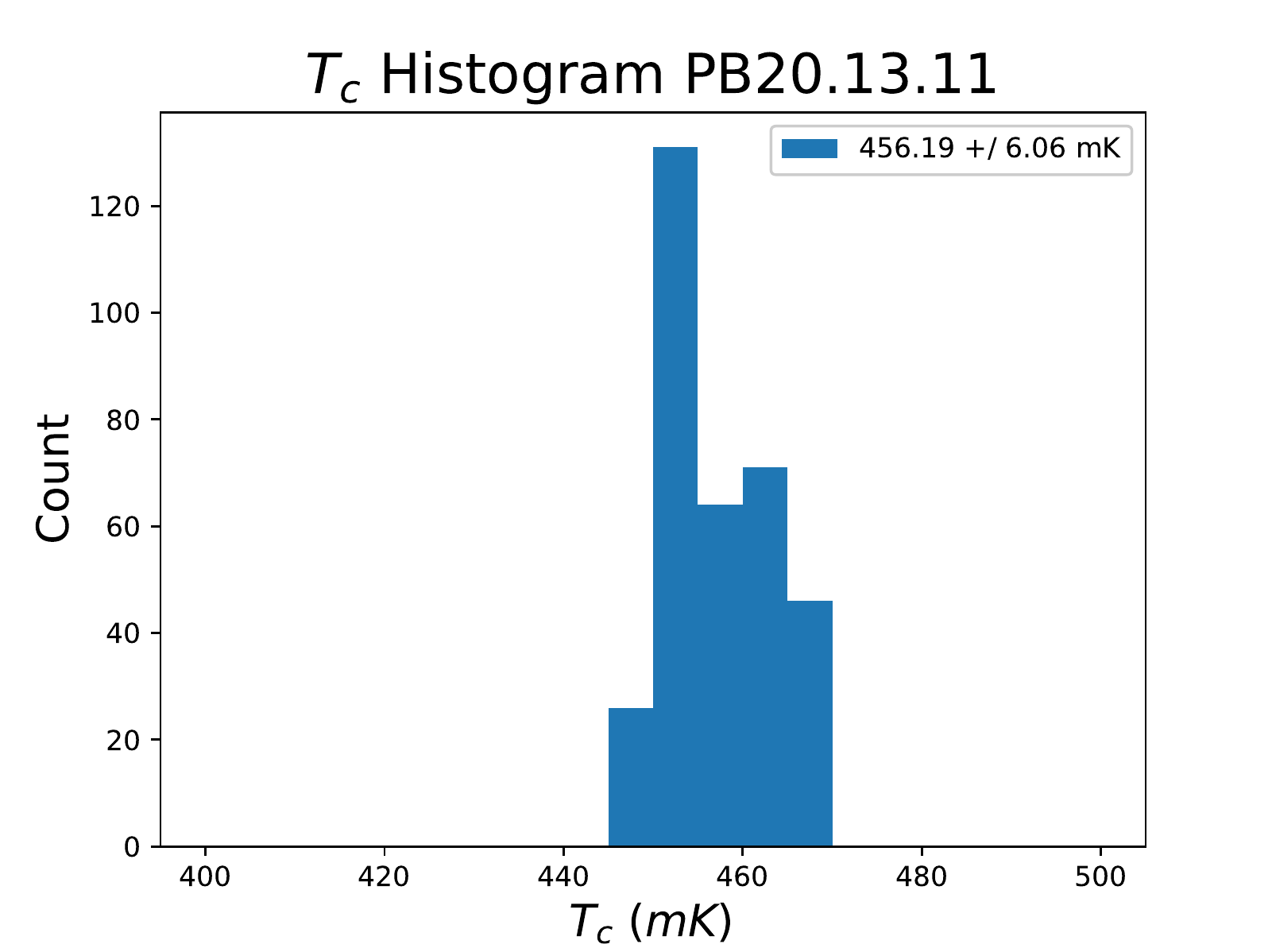}\\ 
\hline
\hline
& & \\
$T_c$ ($mK$)&  \includegraphics[width=0.3\textwidth]{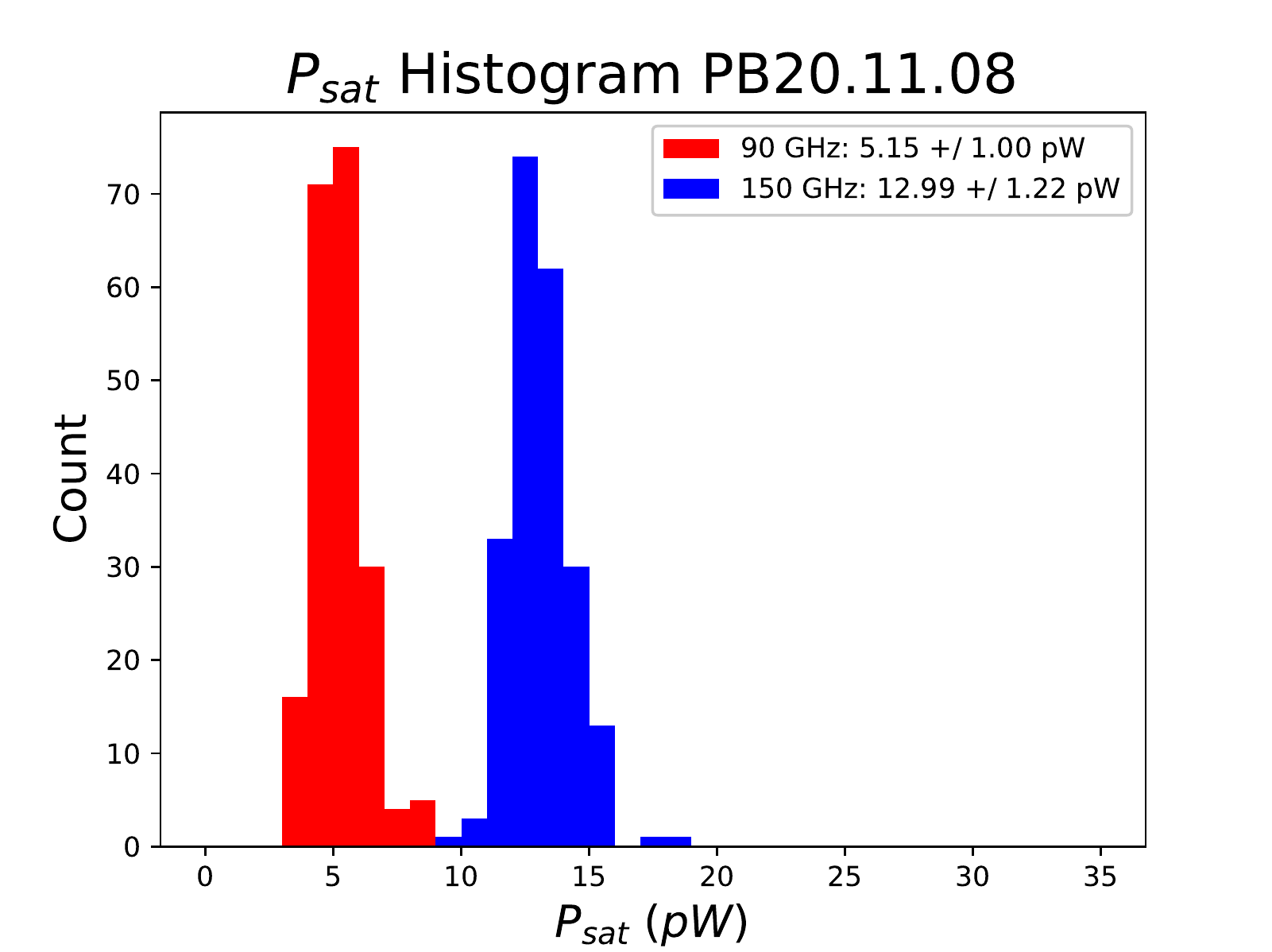} &  \includegraphics[width=0.3\textwidth]{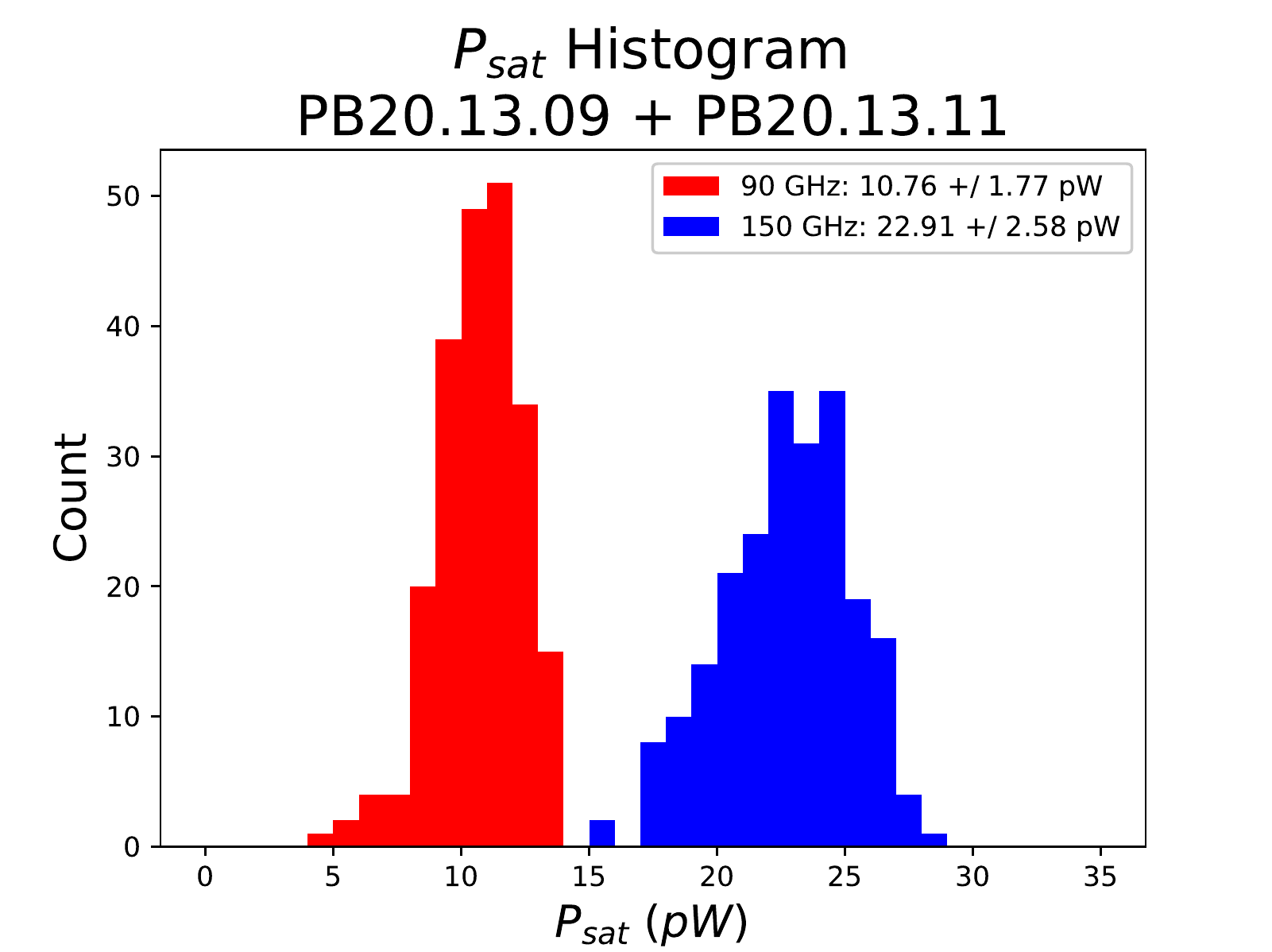}\\ 
\hline
\end{tabular}
\caption{Histograms showing distributions for $R_n$, $T_c$, and $P_{sat}$ from both V11 and V13 arrays.  The targets for $R_n$ and $T_c$ are the same betwen V11 and V13.  The target for V13 $P_{sat}$ is higher by a factor of $\sim$~2 after accounting for the boost in optical efficiency. }\label{table:Dark_Characterization}
\end{center}
\end{figure}

\vspace{-10pt}
In this paper, we only report on the results of dark characterization and refer the reader to Elleflot 2018 for a detailed description of bolometer characterization procedures and algorithms.\cite{Elleflot_PB2Readout_2018}  We present respresntative historgrams of $R_n$, $T_c$, and $P_{sat}$ from both V11 and V13 arrays in Table~\ref{table:Dark_Characterization}.

\vspace{-15pt}
\subsubsection{Optical Characteriztation}

\vspace{-10pt}
We present the optical characterization for a single array characterized at UC Berkeley.  For this test we raised the Tc of the TES's of one of our V13 arrays in order to make the bolometer have saturation powers suitable for laboratory measurements.   We optically characterized 4 bolometers in 3 pixels to verify the optical performance of the V13 arrays. 

\begin{figure}[htbp]
   \centering
   \subfigure[PB2-AB Bands]
  {\includegraphics[width=0.8\textwidth]{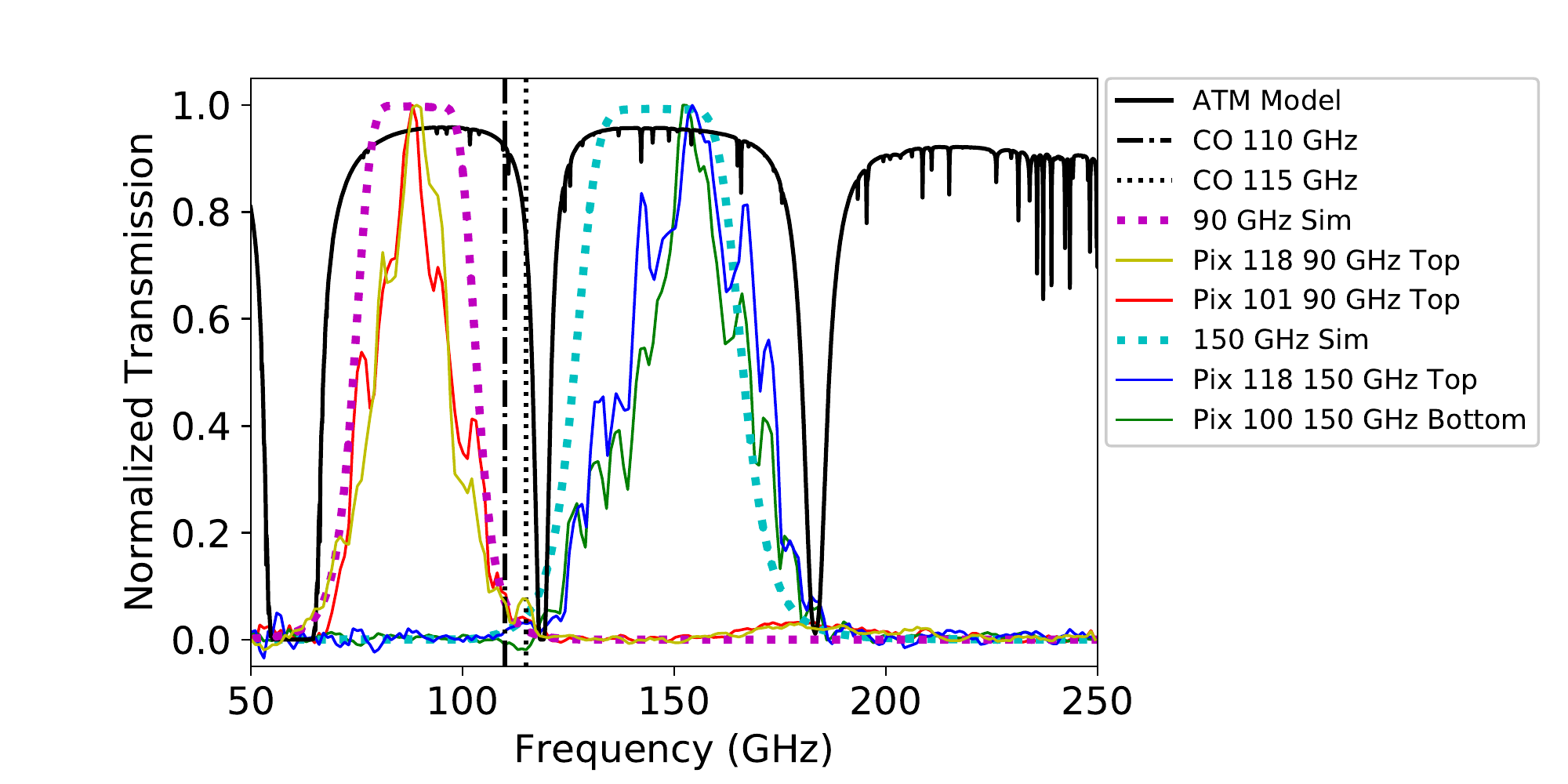}\label{subfig:PB2AB_Spectra}}
   \subfigure[PB2-AB Bands]
  {\includegraphics[width=0.5\textwidth]{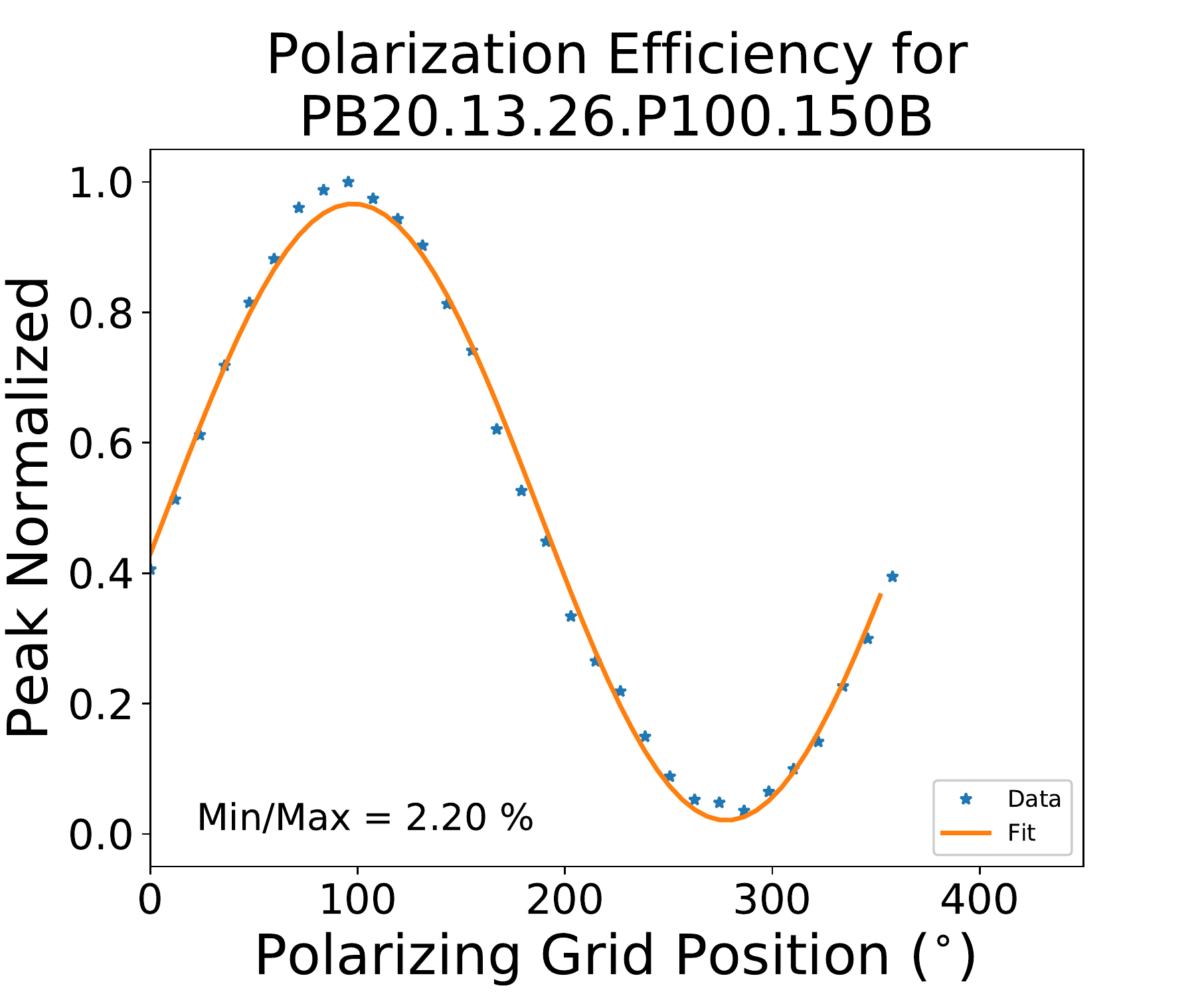}\label{subfig:PB2AB_PolEff}}
   \subfigure[90GHz Beam]
  {\includegraphics[width=0.39\textwidth]{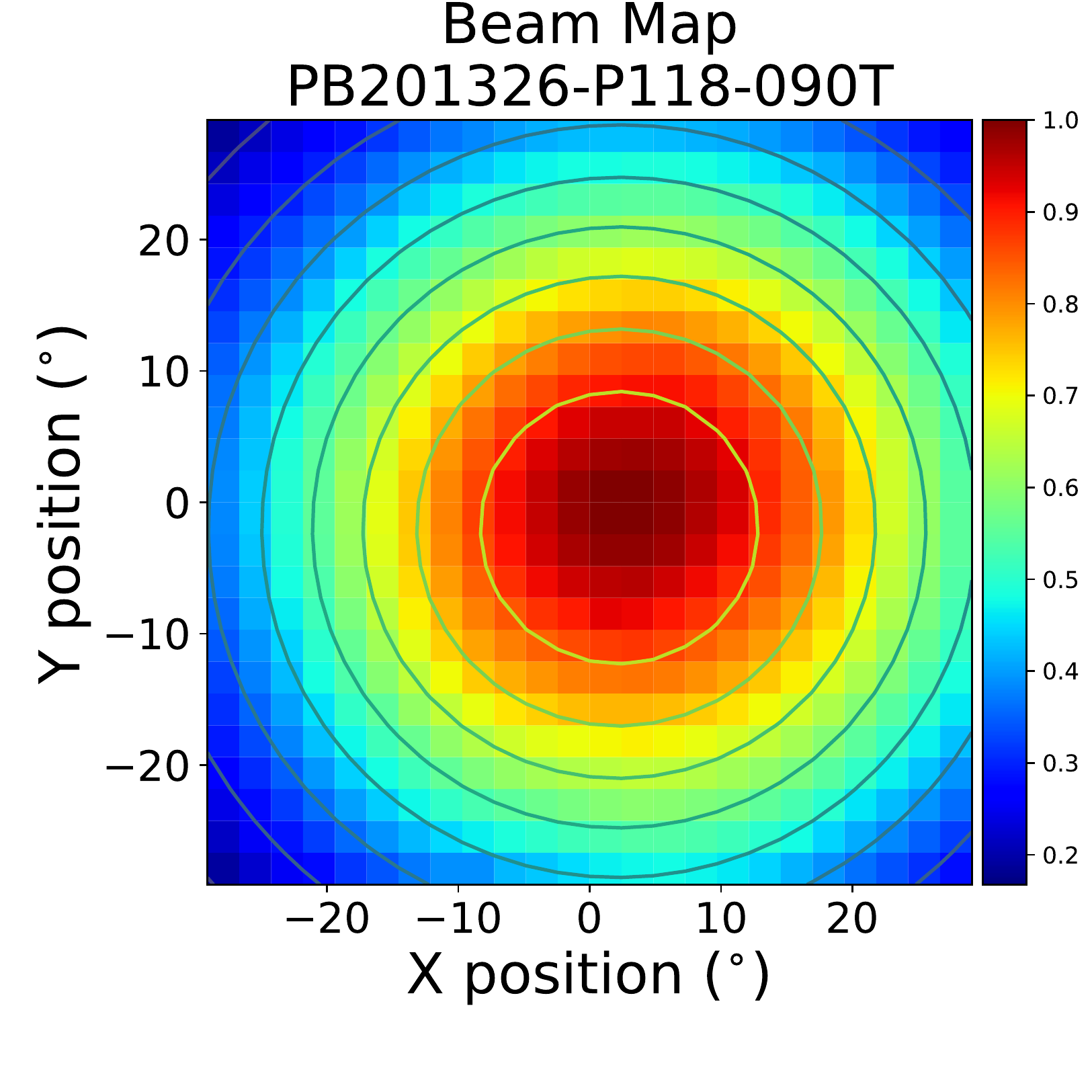} \label{subfig:PB2AB_90_Beam}}
\caption{Peak normalized spectra from array PB20-13-26 are shown in \ref{subfig:PB2AB_Spectra}.   The spectra match the simulated bands (dash lines) and fit intow the atmospheric windows (solid black line) at the POLARBEAR site.  They also avoid the carbon monoxide lines at 110 (dot dash line) and 115 GHz (dotted line).  Each band has a bolometer for each frequency and polarization.  A plot of a bolometers response to a rotating linearly polarized grid (blue stars) and fit (yellow line) is shown in \ref{subfig:PB2AB_PolEff}.  A beammap of a 90 GHz bolometer taken from array PB2-13-26 using a lenslet nearly identical to those we plan to deploy is shown in subfigure \ref{subfig:PB2AB_90_Beam}.  The data are shown as the color map and the contours are two dimensional gaussion fits to the data.}\label{fig:optical_characterization}
\end{figure}

We characterized the band passes, beams, and polarization efficiency of these pixels and present a subset of this data in Fig.~\ref{fig:optical_characterization}.  We find that the pixels have beam ellipticities of 2\% and 8\%,  polarization efficiency of 95.4\% and 97.8\%, and bandwidths of 89 $\pm$ 12.5 GHz and 148 $\pm$ 20 GHz for the 90 and 150 bands respectively.

\vspace{-20pt}
\subsection{Conclusion}

\vspace{-10pt}
We plan to deploy PB2-A, the first of three receivers in the Simons Array in 2018.  We have fabricated enough candidate V11 (SiOx with ``cross overs'') and V13 (SiNx with ``cross unders'') arrays to populate PB2-A focal plane with either version.   Array characterization is underway to determine the final focal plane configuration.  Initial testing results of these candidate arrays indicate that we have both good inter and intra array uniformity. The V13 arrays have higher yield and better optical efficiency than V11 arrays and are more likely to be deployed.   

\vspace{-5pt}
\begin{acknowledgements}

\vspace{-10pt}
We acknowledge support from the MEXT Kahenhi grant 21111002, NSF grant AST-0618398, NASA grant NNG06GJ08G, The Simons Foundation, Natural Sciences and Engineering Research Council, Canadian Institute for Advanced Research, Japan Society for the Promotion of Science, and the CONICYT provided invaluable funding and support. Detectors were fabricated at the Berkeley Marvell Nanofabrication laboratory.

\include{references.bib}

\end{acknowledgements}

\bibliographystyle{unsrt}
\bibliography{bibliography.bib}

\end{document}